\documentclass[aps,prb,preprint,groupedaddress]{revtex4}

\usepackage{graphicx}
\usepackage{dcolumn}
\usepackage{bm}
\usepackage{amsmath,amssymb}

\begin{document}


\title{Spin Nematic Liquids of the S=1 Spin Ladder in Magnetic Field}



\author{T\^oru \textsc{Sakai}$^{1,2}$ and Kiyomi \textsc{Okamoto}$^{1}$}
\affiliation{
$^{1}$Graduate School of Material Science, University of Hyogo, Hyogo 678-1297, Japan \\
$^{2}$National Institutes for Quantum and Radiological Science and Technology (QST), SPring-8, Hyogo 679-5148, Japan
}


\date{Received September 1, 2019}

\begin{abstract}

The magnetization process of the $S=1$ spin ladder system is 
investigated using the numerical exact diagonalization of 
finite-size clusters. 
The field-induced spin nematic liquid phase was predicted to 
appear by our previous work. 
Several ground-state phase diagrams in the plane of the 
single-ion anisotropy and the external magnetic field are obtained 
in the present study. 
\end{abstract}


\maketitle


\section{Introduction}

The spin nematic phase is one of interesting topics in the 
field of the strongly correlated electron systems. 
For example, the high-magnetic field measurement of 
the quasi-one-dimensional compound LiCuVO$_4$ detected 
it\cite{nawa}. 
In addition the spin-liquid-like behavior of the 
$S=1$ triangular-lattice compound NiGa$_2$S$_4$ 
was theoretically explained by the spin nematic phase\cite{nakatsuji}. 
The spin nematic order is the quadrapole order of the quantum spins in 
two- or three-dimensional systems. 
On the other hand it appears as the gapless Tomonaga-Luttinger 
liquid phase of the two-magnon bound state in one-dimensional 
systems. 
In our previous work\cite{yamanokuchi} 
using the numerical exact diagonalization 
of finite-clusters the spin nematic Tomonaga-Luttinger liquid (TLL) phase 
was revealed to occur in the $S=1$ spin ladder system under the 
external magnetic field in the presence of sufficiently large 
negative single-ion anisotropy. 
In addition several ground-state phase diagrams in the plane of 
the anisotropy and the magnetization were presented. 
However, in order to propose some experiments to detect the 
spin nematic TLL phase in real materials, 
the phase diagram in the plane of the anisotropy and 
the external magnetic field 
would be much more useful. 
In this paper, we investigate the $S=1$ spin ladder system 
with the numerical exact diagonalization of finite-size 
clusters and obtain the phase diagrams in the anisotropy and field 
plane. 

\section{Model}

The $S=1$ spin ladder with the single-ion anisotropy $D$ is described by 
the Hamiltonian 
\begin{eqnarray}
{\cal H}=
   &&J_1\sum _{i=1,2} \sum_{j=1}^L \vec{S}_{i,j}\cdot \vec{S}_{i,j+1} 
    +J_{\rm r} \sum_{j=1}^L \vec{S}_{1,j}\cdot \vec{S}_{2,j}   \nonumber \\
   && +D \sum_{i=1,2}\sum_{j=1}^L(S^z_{i,j})^2  \nonumber \\
   && -H \sum_{i=1,2}\sum_{j=1}^LS^z_{i,j},
\end{eqnarray}
where $\vec{S}_{i,j} = (S_{i,j}^x,S_{i,j}^y,S_{i,j}^z)$ denotes the spin-1 operator acting on the spin
at the $j$th rung and the $i$th chain. 
The quantity $J_1$ denotes the nearest neighbor leg interaction constant,
$J_{\rm r}$ the rung interaction constant,
and $H$ the strength of the external 
magnetic field along the $z$ direction. 
We investigate the ground state of this model using the numerical 
exact diagonalization of finite-size cluster up to $L=8$. 
Throughout this paper we consider the negative $D$ only, namely 
the easy-axis anisotropy, 
and fix $J_1=1.0$. 

\section{Ground state under $H=0$}

\begin{figure}[h]
\bigskip
      \centerline{\includegraphics[scale=0.4]{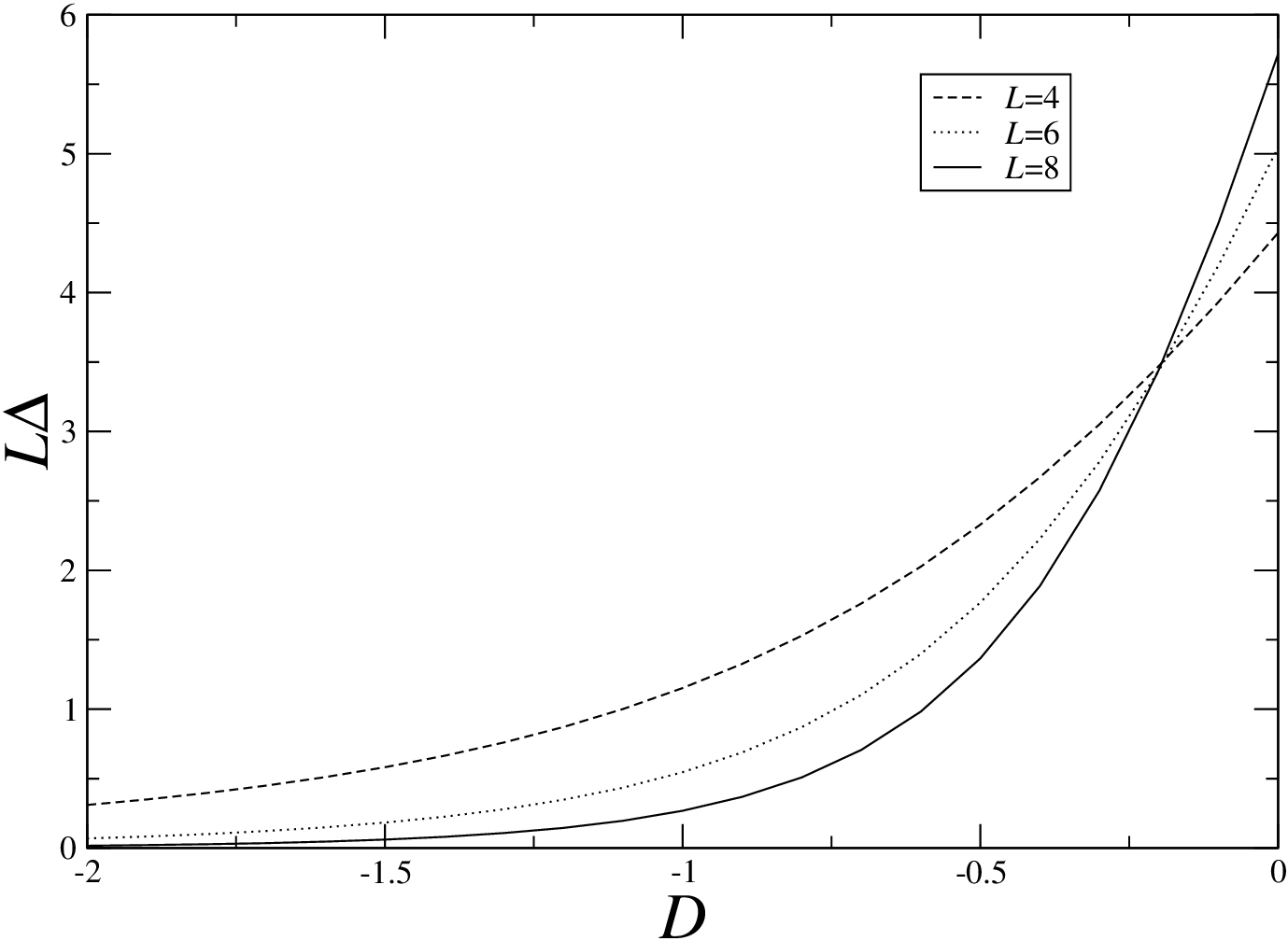}}
  \caption{Scaled gap $L\Delta_L(D)$ plotted versus $D$ for 
  $J_1=J_{\rm r}=$2.0. Since the scaled gaps for $L=$4, 6 and 8 {cross} 
  to each other almost at the same point, we use $D_{{\rm c},6}=-0.20$ 
  as the best estimation of the quantum critical point $D_{\rm c}$. }
  \label{prg20}
\end{figure}

In the absence of the external magnetic field, 
{for $D=0$,
the system 
is in the plaquette singlet state
which is non-degenerate and has the spin gap
\cite{todo2001}. 
On the other hand, for sufficiently large negative $D$, 
the system is in the N\'eel state along the $z$-direction
which is doubly degenerate and also has the spin gap.}
The critical point $D _{\rm c}$ can be estimated 
by the  phenomenological renormalization group method. 
Namely, the size-dependent critical point 
$D_{{\rm c},L}$ is determined from the equation for the scaled gaps 
\begin{eqnarray}
(L+2)\Delta_{L+2}(D_{{\rm c} ,L})=L\Delta_{L}(D_{{\rm c},L}),
\end{eqnarray}
where $\Delta_L(D)$ is 
the lowest energy gap between the $k=0$ ground state and 
the $k=\pi$ subspace in the leg direction.
The scaled gap $L\Delta_L(D)$ is plotted versus {$D$} for 
$J_1=J_{\rm r}=2.0$ in Fig. \ref{prg20}. 
Since the size dependence of $D_{{\rm c},L}$ is quite small, 
we use $D_{{\rm c},6}=-0.20$ as the best estimation of the critical point $D_{\rm c}$. 

\section{Tomonaga-Luttinger liquid phases for $H>0$}

Since the ground state {at $H=0$} is in the plaquette singlet phase 
for $0> D > D_{\rm c}$, 
a phase transition occurs at the critical field $H_{{\rm c}1}$ and the 
gapless TLL phase is realized for $H>H_{{\rm c}1}$\cite{sakai1991a,sakai1991b}. 
On the other hand, {when the ground state is the N\'eel ordered state for $D < D_{\rm c}$,}
the magnetization process is expected to be 
similar to that of the $S=1/2$ Ising-like $XXZ$ ladder. 
In this case the TLL phase is also realized 
{above} the critical field $H_{{\rm c}1}$. 
The quasiparticle excitation, however, is different between these 
two TLL phases. 
Each elementary magnon excitation should occur by $\delta S^z=2$, 
because the $S^z=0$ state cannot occur for sufficiently large negative $D$ 
($D<D_{\rm c}$), while $\delta S^z=1$ for $D > D_{\rm c}$. 
The former TLL phase is called the spin nematic TLL
phase, to distinguish from the latter one, namely, the conventional TLL phase. 
These two TLL phases can be distinguished by whether the gapless 
excitation is $\delta S^z=1$ or 2. 

\section{Phase diagrams on the $D$-$H$ plane}

The purpose of this paper is to obtain the ground-state phase diagram 
on the $D$-$H$ plane. 
We define $H_{\rm c1}$ as the critical field where the non-zero magnetization 
appears for the first time with increasing $H$. 
At first, we should consider the possibility of the magnetization jump 
at $H_{\rm c1}$. 
The field $H_{\rm jump}(M)$ is defined as 
\begin{eqnarray}
H_{\rm jump}(M)=[E(M)-E(0)]/M,
\label{jump}
\end{eqnarray}
where $E(M)$ is the lowest energy for $\sum _jS^z_j=M$. 
When the magnetization jump occurs to $M$ at $H_{\rm c1}$, 
$H_{\rm jump}(M) < H_{\rm jump}(M')$ is satisfied for 
every $M'$ ($M'<M$) and $H_{\rm c1}$ corresponds to $H_{\rm jump}(M)$. 
The present numerical diagonalization for $L=8$ indicates that 
the smallest $H_{\rm jump}(M)$ is given for $M \geq 3$ in the whole region 
$D<D_c$. 
Thus the $M=2$ state is skipped and the magnetization jump occurs 
at $H_{\rm c1}$. 

In order to estimate the phase boundary between the two TLL phase 
in the finite magnetization phase, the cross points between the 
$\delta S^z =1$ excitation gap and the $2k_{\rm F}$ excitation 
gap of the two magnon bound state in our previous work. 
In this paper, however, we use the cross points between 
$H_1(M)$ and $H_2(M)$ defined as 
\begin{eqnarray}
H_1(M)&=&E(M+1)-E(M) \\
H_2(M)&=&[E(M+2)-E(M)]/2,
\label{h1h2}
\end{eqnarray}
for $L=8$, because more points can be obtained than the previous method. 

The saturation field $H_{\rm sat}$ is obtained as 
$E(2L)-E(2L-1)$ for larger $D$, while $[E(2L)-E(2L-2)]/2$ for smaller $D$. 

\begin{figure}[h]
  \centerline{\includegraphics[scale=0.4]{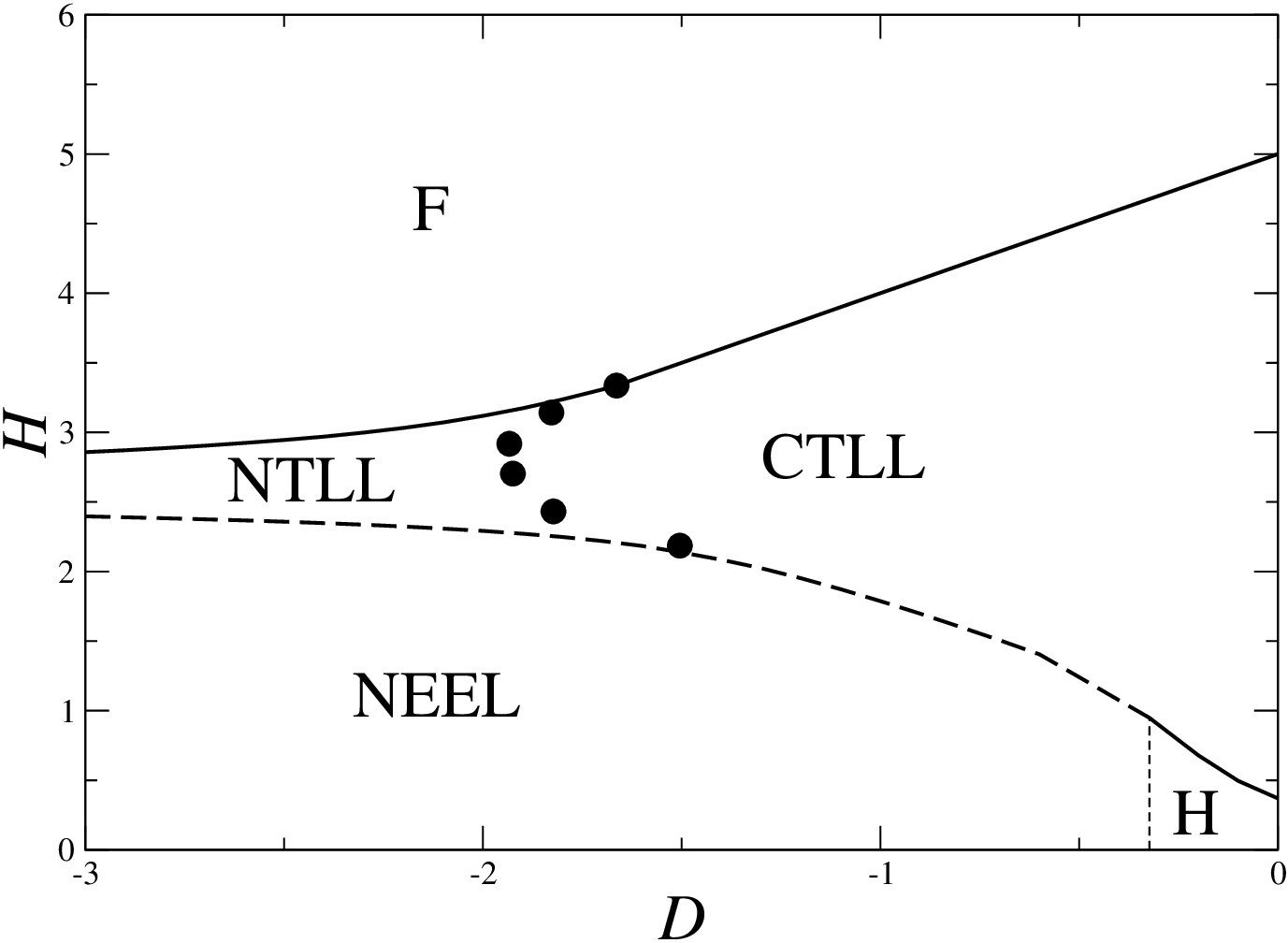}}
  \caption{Phase diagram on the $D$-$H$ plane for $J_1=1.0$ and 
  $J_{\rm r}=0.5$. 
  Here H, NEEL, CTLL, NTLL and F correspond to the Haldane, the N\'eel, 
  the conventional Tomonaga-Luttinger liquid, 
  the nematic Tomonaga-Luttinger liquid, and the ferromagnetic phases, 
  respectively. 
  Dashed curve is the critical magnetic field with the magnetization jump.
  }
  \label{jr05}
  \bigskip
\end{figure}

\begin{figure}[ht]
  \centerline{\includegraphics[scale=0.4]{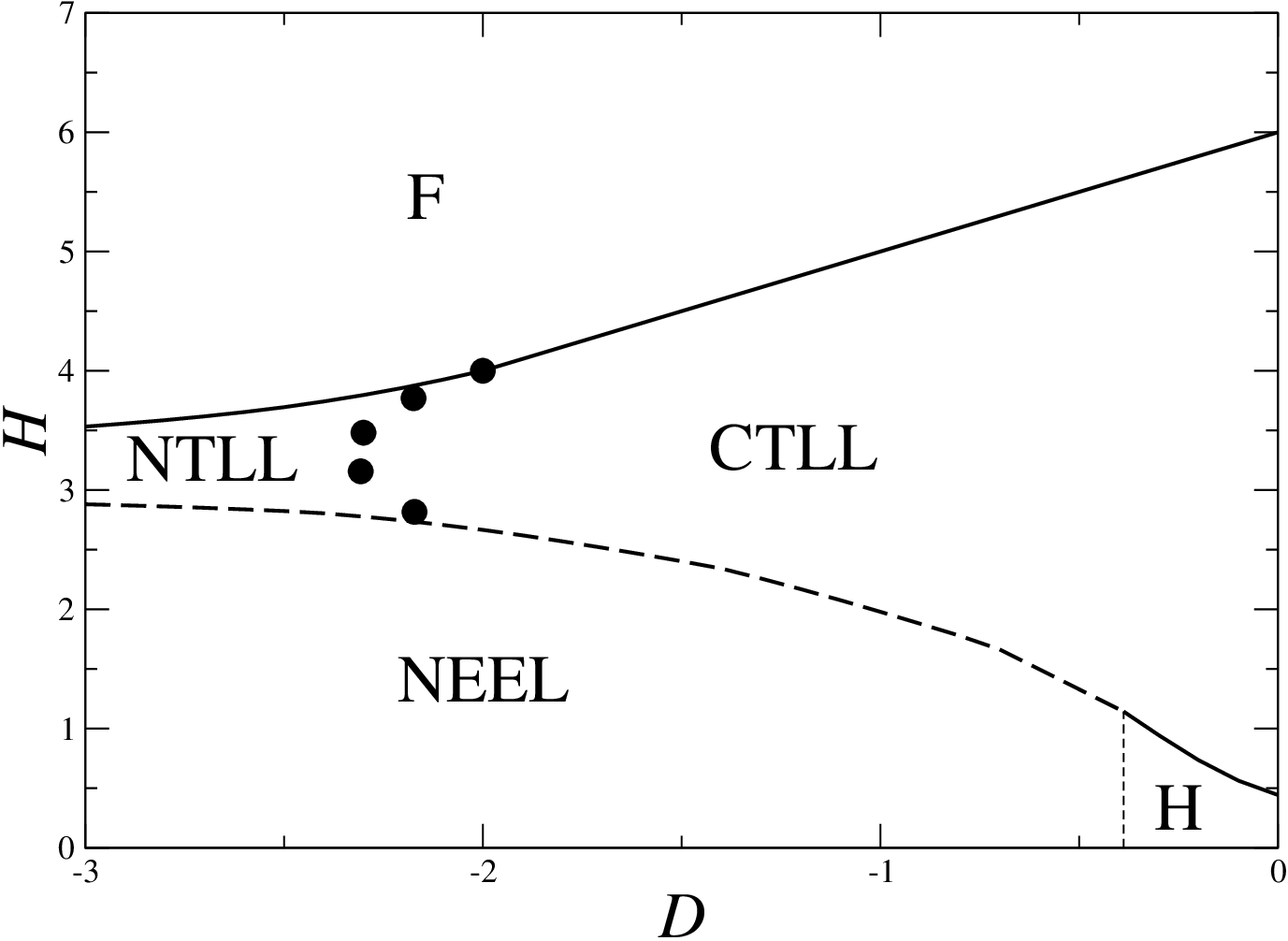}}
  \caption{Phase diagram on the $D$-$H$ plane for $J_1=J_{\rm r}=1.0$. 
 }
 \label{jr10}
\end{figure}

\begin{figure}[h]
  \centerline{\includegraphics[scale=0.4]{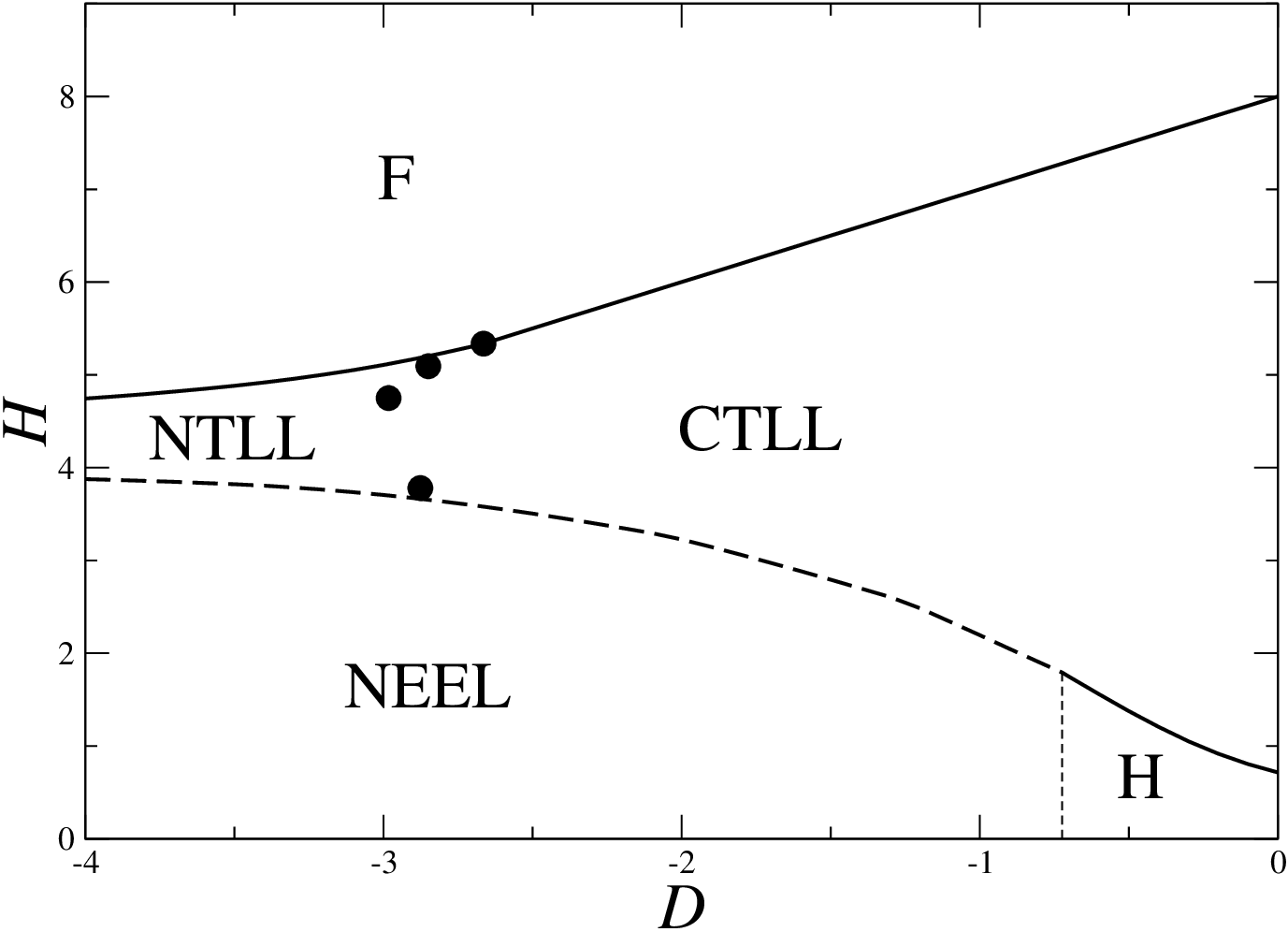}}
  \caption{Phase diagram on the $D$-$H$ plane for $J_1=1.0$ and 
  $J_{\rm r}=2.0$. 
  }
  \label{jr20}
\end{figure}

Using the numerical diagonalization for $L=8$, we obtain the $H$-$D$ 
phase diagrams for $J_r=$0.5, 1.0 and 2.0 shown in Figs. \ref{jr05}, 
\ref{jr10} and \ref{jr20}, respectively. 
In our previous work the system size dependence of these phase boundaries 
is quite small. 
In these phase diagrams the Haldane (H), the N\'eel ordered (NEEL), 
the conventional Tomonaga-Luttinger liquid (CTLL), 
the nematic Tomonaga-Luttinger liquid (NTLL), and 
the ferromagnetic (F) phases appear. 
Dashed curves are the critical magnetic field with the magnetization jump. 

\begin{figure}[h]
  \bigskip
  \centerline{\includegraphics[scale=0.4]{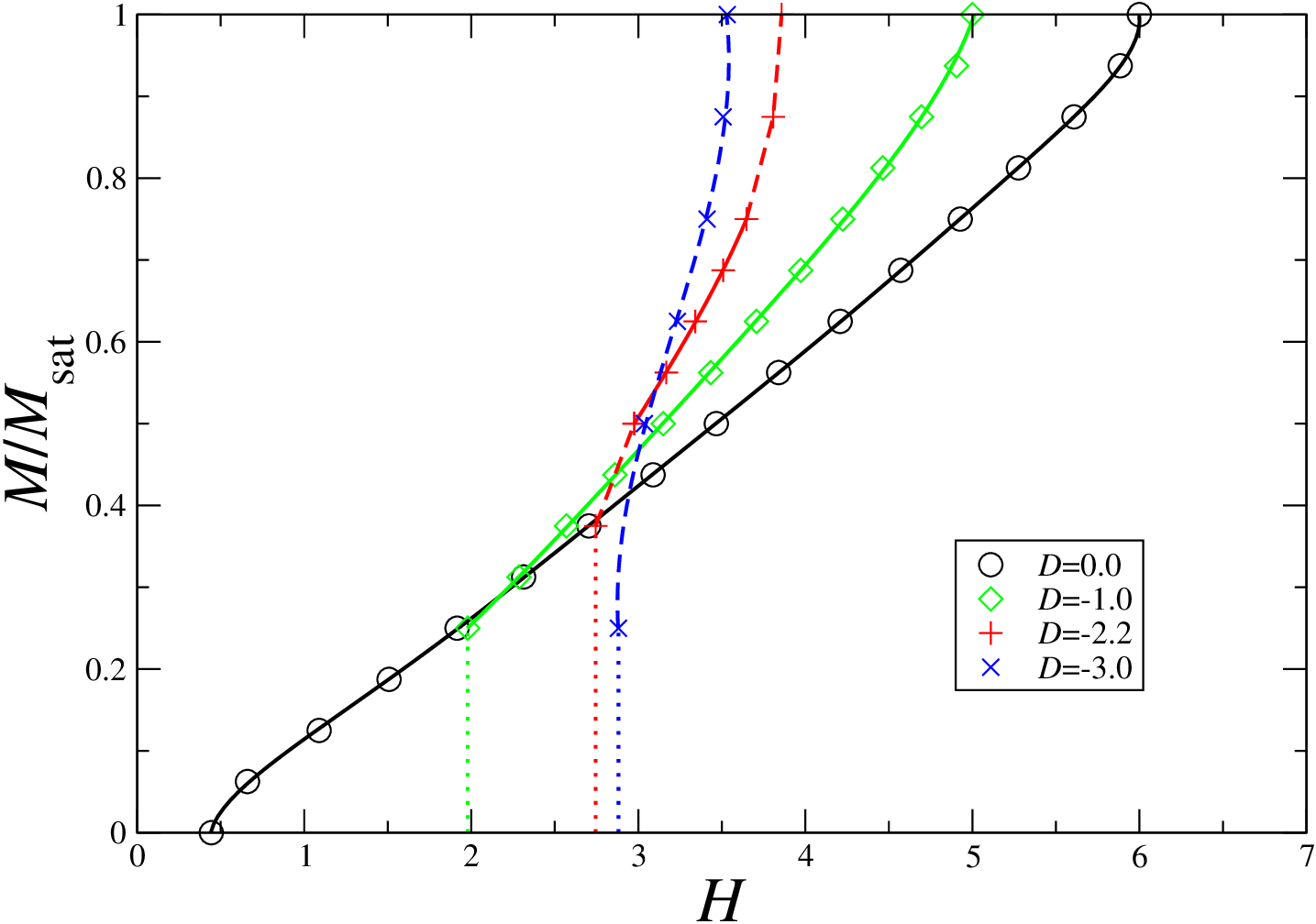}}
  \caption{Ground state magnetization curves obtained from the 
  numerical diagonalization for $L=8$ in the case of $J_1=J_{\rm r}=1.0$. 
  Black, green, red and blue symbols 
correspond to $D=$0, $-1.0$, $-2.2$ and $-3.0$, respectively. 
Lines and curves are guides for the eye. 
Solid and dashed curves correspond to the conventional and the 
nematic TLL phases, and dotted lines mean the magnetization jump. 
  }
  \label{mag}
\end{figure}

\section{Magnetization curve}

In order to consider the experiment to detect the nematic TLL phase, 
it would be useful to give the theoretical magnetization curve 
based on the numerical diagonalization for $L=8$. 
If the magnetization $M$ is realized for $H_-(M)<H<H_+(M)$ in the 
ground state of the finite-size system, 
the averaged field $H_{\rm av}(M)=[H_-(M)+H_+(M)]/2$ is used 
to obtain the magnetization curve in the present work. 
Namely, we neglect the finite-size correction proportional to 
$1/L^2$ here. 
In the case of $J_1=J_{\rm r}=1.0$, the ground state magnetization 
curves are 
shown in Fig. \ref{mag}, where black, green, red and blue symbols 
correspond to $D=$0, $-1.0$, $-2.2$ and $-3.0$, respectively. 
Lines and curves are guides for the eye. 
Solid and dashed curves correspond to the conventional and the 
nematic TLL phases, and dotted lines mean the magnetization jump. 
The critical points between the two TLL phases are not so precise, 
because they still include some finite-size effects. 
It suggests that the reentrant quantum phase transition can 
occur for $D=-2.2$.

\section{Summary}
The $S=1$ spin ladder with the easy-axis single-ion anisotropy under the 
magnetic field is investigated using the numerical exact diagonalization 
of finite-size clusters. 
We obtain the ground-state phase diagrams in the $D$-$H$ plane 
including the conventional and nematic TLL phases and 
the magnetization jump. Some magnetization curves are also obtained.

\section*{Acknowledgment}
This work was partly supported by JSPS KAKENHI, Grant Numbers 16K05419, 
16H01080 (J-Physics) and 18H04330 (J-Physics). 
A part of the computations was performed using 
facilities of the Supercomputer Center, 
Institute for Solid State Physics, University of Tokyo, 
and the Computer Room, Yukawa Institute for Theoretical Physics, 
Kyoto University.

\end{document}